\documentclass[letter]{aa} % for the letters 
\usepackage{graphicx}

\usepackage{lscape}
\usepackage{amssymb}
\usepackage{amsmath}

\usepackage{txfonts}

\usepackage{booktabs}
\usepackage{subfigure}
\usepackage{placeins}
\usepackage{verbatim}
\usepackage{adjustbox}
\usepackage[toc,page]{appendix} 
\usepackage{tabularx}
\usepackage{lscape}
\usepackage{textcomp}
\usepackage{gensymb}
\usepackage{multicol}
\usepackage[symbol]{footmisc}
\usepackage{tikz}
\usepackage{caption}
\usepackage{longtable}

\usepackage{natbib}
\usepackage{supertabular}

\usepackage{color}
\usepackage{lscape}

\usepackage{hyperref}  
\hypersetup{colorlinks=true,linkcolor=[rgb]{1.,0.2,0.2},citecolor=[rgb]{0.1,0.4,1.},filecolor=[rgb]{0.7,0.2,0.2},urlcolor=[rgb]{0.7,0.2,0.2}}

\newcommand{\bd}{\begin{displaymath}}
\newcommand{\ed}{\end{displaymath}}
\newcommand{\be}{\begin{equation}}
\newcommand{\ee}{\end{equation}}
\newcommand{\beaa}{\begin{eqnarray*}}
\newcommand{\eeaa}{\end{eqnarray*}}
\newcommand{\bea}{\begin{eqnarray}}
\newcommand{\eea}{\end{eqnarray}}

\def\macs1149{MACS 1149}

\begin{document} 

   \title{Augmenting the power of time-delay cosmography in  lens galaxy clusters by probing their member galaxies. II. Cosmic chronometers}

   %\subtitle{}

\author{
P.~Bergamini \inst{\ref{unimi}, \ref{inafbo}} 
\fnmsep\thanks{E-mail: \href{mailto:pietro.bergamini@unimi.it}{pietro.bergamini@unimi.it}}
\and
S.~Schuldt \inst{\ref{unimi},\ref{inafmilano}} \and
A.~Acebron \inst{\ref{unimi},\ref{inafmilano}} 
\and
C.~Grillo \inst{\ref{unimi},\ref{inafmilano}} \and
U.~Me\v{s}tri\'{c} \inst{\ref{unimi}, \ref{inafbo}}\and
G.~Granata \inst{\ref{unimi},\ref{inafmilano}} \and
G.~B.~Caminha \inst{\ref{tum}, \ref{mpa}}\and
M.~Meneghetti \inst{\ref{inafbo}} \and
A.~Mercurio \inst{\ref{salerno}, \ref{inafna}, \ref{salernoINFN}} \and
P.~Rosati \inst{\ref{unife},\ref{inafbo}} \and
S.~H.~Suyu \inst{\ref{tum}, \ref{mpa}, \ref{asiaa}}\and
E.~Vanzella \inst{\ref{inafbo}}
}

\institute{
Dipartimento di Fisica, Universit\`a  degli Studi di Milano, via Celoria 16, I-20133 Milano, Italy \label{unimi}
\and
INAF -- OAS, Osservatorio di Astrofisica e Scienza dello Spazio di Bologna, via Gobetti 93/3, I-40129 Bologna, Italy \label{inafbo} 
\and
INAF -- IASF Milano, via A. Corti 12, I-20133 Milano, Italy \label{inafmilano}
\and
Technical University of Munich, TUM School of Natural Sciences, Department of Physics, James-Franck-Str 1, 85748 Garching, Germany \label{tum}
\and
Max-Planck-Institut f\"ur Astrophysik, Karl-Schwarzschild-Str. 1, D-85748 Garching, Germany \label{mpa}
\and
Università di Salerno, Dipartimento di Fisica "E.R. Caianiello", Via Giovanni Paolo II 132, I-84084 Fisciano (SA), Italy \label{salerno}
\and
INAF -- Osservatorio Astronomico di Capodimonte, Via Moiariello 16, I-80131 Napoli, Italy \label{inafna}
\and
INFN – Gruppo Collegato di Salerno - Sezione di Napoli,  Dipartimento di Fisica "E.R. Caianiello", Università di Salerno, via Giovanni Paolo II, 132 - I-84084 Fisciano (SA), Italy. \label{salernoINFN}
\and
Dipartimento di Fisica e Scienze della Terra, Universit\`a degli Studi di Ferrara, via Saragat 1, I-44122 Ferrara, Italy \label{unife}
\and
Academia Sinica Institute of Astronomy and Astrophysics (ASIAA), 11F of ASMAB, No.1, Section 4, Roosevelt Road, Taipei 10617, Taiwan \label{asiaa}
}   

   %\date{Received September 15, 1996; accepted March 16, 1997}

% \abstract{}{}{}{}{} 
% 5 {} token are mandatory

 \abstract{We present a novel approach to measuring the expansion rate and the geometry of the Universe, which combine time-delay cosmography in lens galaxy clusters with pure samples of 'cosmic chronometers' (CCs) by probing the member galaxies. 
 The former makes use of the measured time delays between the multiple images of time-varying sources strongly lensed by galaxy clusters, while the latter exploits the most massive and passive cluster member galaxies to measure the differential time evolution of the Universe. 
 We applied two different statistical techniques, adopting realistic errors on the measured quantities, to assess the accuracy and the gain in precision on the values of the cosmological parameters. 
 We demonstrate that the proposed combined method allows for a robust and accurate measurement of the value of the Hubble constant. In addition, this provides valuable information on the other cosmological parameters thanks to the complementarity between the two different probes in breaking parameter degeneracies.
 Finally, we showcase the immediate observational feasibility of the proposed joint method by taking advantage of the existing high-quality spectro-photometric data for several lens galaxy clusters. 
 }

   
   \titlerunning{Combining TDC with cosmic chronometers in lens galaxy clusters}
   \authorrunning{Bergamini et al.}
   \maketitle

%
%-------------------------------------------------------------------

\section{Introduction}
\label{sec:intro}

Emergent, independent probes for measuring the present-day expansion rate, defined as the Hubble constant ($H_0$), and the geometry of the Universe can offer valuable insights into unknown systematic effects and help clarify current tensions in cosmology \citep[][]{Verde2019, Moresco2022}.
Time-delay cosmography (TDC) grants such an opportunity. Almost sixty years after \citet{Refsdal1964}'s original idea that strongly lensed supernovae (SNe) with measured time delays between their multiple images could offer a novel way to measure the value of $H_0$, this technique has provided competitive estimates of its value \citep[e.g.][]{Suyu2017, Treu2016b, Grillo2018, Birrer2019, Wong2020, Shajib2023}.

Using time-varying sources strongly lensed by galaxy clusters represents a complementary and powerful avenue.
The multiply-imaged SN `Refsdal' in the galaxy cluster MACS~J1149.5$+$2223 \citep[hereafter, \macs1149, e.g.][]{Lotz2017} has allowed for the first measurement of the value of $H_0$ via a multiply-imaged SN with measured time delays \citep{Kelly2016, Kelly2023}. In particular,  \citet{Grillo2018, Grillo2020}, using a full strong lensing analysis, inferred the value of $H_0$ with a 6\% total (statistical plus systematic) uncertainty.
In this series of two letters, we present novel methods to enhance and complement the power of TDC in lens galaxy clusters, first by observing Type Ia Supernovae (SNe Ia) in cluster member galaxies \citep[][Letter 1, hereafter]{Acebron2023} and then by using a subset of member galaxies in lens galaxy clusters as 'cosmic chronometers' (CCs).

As shown in \cite{Grillo2018, Grillo2020}, to take full advantage of lens clusters as cosmological probes, it is necessary to construct an accurate cluster total mass model by leveraging extensive high-quality spectro-photometric datasets \citep[e.g.][]{Caminha2017, Bergamini2019, Lagattuta2023, Acebron2022}. In parallel, these data (already available for several lens galaxy clusters) enable us to homogeneously select pure samples of red, massive, and passive cluster members and exploit them as CCs \citep{Jimenez2002, Stern2010a}.
In this work, we explore, for the first time, the idea of taking advantage of some member galaxies as CCs to yield an independent measurement of the value of the Hubble parameter at the lens cluster redshift. From the combination of these two cosmological probes, we can then assess the gain in precision on the measurement of the values of the main cosmological parameters.

The letter is organised as follows. In Sect.~\ref{sec:methods}, we concisely illustrate the principles of the TDC and CC probes to infer the values of the cosmological parameters. In Sect.~\ref{sec:sims}, we detail our methods to quantify the precision attainable in these measurements with the proposed joint technique. In Sect.~\ref{sec:discussion}, we present our results and describe the observational feasibility of carrying out this analysis. Finally, we draw our conclusions in Sect.~\ref{sec:conclusions}.

\section{Methods}
\label{sec:methods}
In this section, we summarise the basic principles behind the TDC and CC techniques used to measure the values of $H_0$, the present-day cosmological densities of matter ($\Omega_{\rm m}$) and dark energy ($\Omega_{\rm de}$), and the dark-energy equation of state parameter ($w$).

\subsection{Time-delay cosmography}

Gravitational lensing refers to the effect according to which the light rays emitted from a background source are deflected by a foreground mass distribution, such as a galaxy or a galaxy cluster. When the total mass density of the gravitational lens is sufficiently high, we reach the strong-lensing regime and lensed sources are multiply imaged.

If the luminosity of a background source is intrinsically variable in time, such as that of quasars or SNe, its different multiple images appear with a delay in time to the observer that can be estimated. In particular, the time delay, 
\begin{equation}
\label{eq:time_delay}
    \Delta t_{\rm i_1 i_2} = \frac{D_{\Delta t}}{c} \Delta \phi_{\rm i_1i_2} ,
\end{equation}
between two images, $\rm i_1$ and $\rm i_2$, of the same background source can be measured through dedicated (photometric) monitoring campaigns \citep[e.g.][]{Courbin2018, Millon2020, Kelly2023}, while the Fermat potential, $\phi$, is obtained by modelling the total mass distribution of the gravitational lens ($c$ is the speed of light). By determining these two quantities, one can probe the so-called time-delay distance, $D_{\Delta t}$ \citep{Suyu2010a}, defined as
\be
\label{eq:time_delay_dist}
D_{\Delta t} = (1+z_\text{d}) \frac{D_\text{d}^\text{A} D_\text{s}^\text{A}}{D_\text{ds}^\text{A}},
\ee
where $z_\text{d}$ is the redshift of the gravitational lens and $D_\text{d}^\text{A}$, $D_\text{s}^\text{A}$, and $D_\text{ds}^\text{A}$ are the angular-diameter distances to the lens, to the source, and between the lens and the source, respectively. From this combination of distances, we see that $D_{\Delta t}$ is proportional to $H_0^{-1}$, with a weaker dependence on the values of the other cosmological parameters ($\Omega_{\rm m}$, $\Omega_{\rm de}$, and $w$). 
The typical relative total (statistical plus systematic) errors on the value of $D_{\Delta t}$ achieved by using a single (galaxy or cluster) gravitational lens range from $\sim4\%$ to $\sim9\%$ \citep[e.g.][]{Suyu2014, Birrer2019, Grillo2020, Wong2020, Shajib2023}.

\subsection{Cosmic chronometers}

The CC technique, introduced by \cite{Jimenez2002}, is designed to probe the cosmic expansion history of the Universe, $H(z)$, independently from the adopted cosmological model. The CC approach exploits a time-redshift relation which, from the definition of $H(z)$ and those of the scale factor and redshift within a Friedmann-Robertson-Walker metric, can be expressed as
\be
\label{eq:hubble_parameterII}
H(z)=-\frac{1}{1+z}\frac{{\rm d} z}{{\rm d} t},
\ee
where ${\rm d} t$ is the differential time evolution of the Universe in a given redshift interval, ${\rm d} z$. The latter can be robustly measured from spectroscopy, while ${\rm d} t$ can be determined by using a homogeneous population of objects, in different redshift bins, as CCs. Here, 
$H(z)$ can be expressed as a function of redshift and of the values of the cosmological parameters, $H(z; H_0, \Omega_{\rm m}, \Omega_{\rm de}, w)$, scaling proportionally with the value of $H_0$ and with a weaker dependence on the other cosmological parameters. 

Extremely massive and passively evolving galaxies, both in blank and cluster fields, have been identified as optimal CCs \citep[e.g.][]{Stern2010a, Moresco2012a, Zhang2014}. 
One of the key steps in this technique is the selection of a pure sample of such CCs, for which a combination of photometric and spectroscopic criteria is commonly adopted \citep[see][for a review]{Moresco2022}. 
In particular, spectroscopy is crucial to ensure the absence of residual emission lines, which can imply the existence of ongoing star formation.
Typically, high stellar-mass or stellar velocity-dispersion cuts are also applied to select the most massive galaxies. The age of the selected CCs can be estimated following different techniques, such as a full spectral-fitting approach \citep[e.g.][]{Simon2005, Stern2010a, Zhang2014, Ratsimbazafy2017}, the analysis of specific features of the galaxy spectrum, known as Lick indices \citep[e.g.][]{Borghi2022a, Borghi2022b}, or of the rest-frame 4000$\,\AA$ break \citep[e.g.][]{Moresco2012a, Moresco2015}.
The total error budget on the measurement of the age of the CCs includes the contributions from several factors, such as residual star-forming contaminants, adopted stellar population synthesis models, and stellar metallicity estimates. With this method, the current precision on the value of $H(z)$ ranges from $\sim10\%$ to $\sim20\%$ \citep[see Table 1 in][]{Moresco2022}.  

\begin{figure*}[ht!]
    \centering
    \includegraphics[width=\columnwidth]{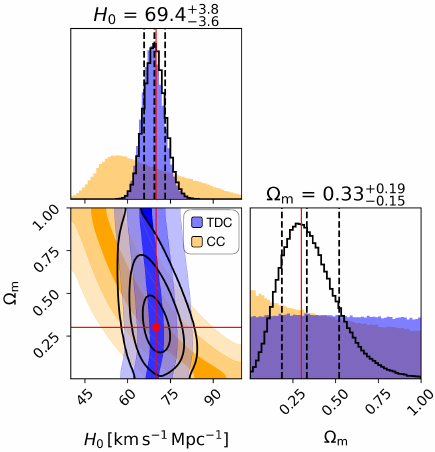}
    \includegraphics[width=\columnwidth]{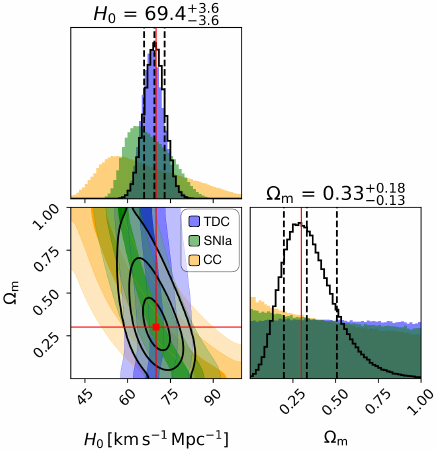} 
    \caption{Inferred values for the cosmological parameters $H_0$ and $\Omega_\text{m}$ in a flat-$\Lambda$CDM model, assuming a total relative uncertainty of 5\%, 10\%, and 5\% for $D_{\Delta t}$, $H(z)$, and $D^\text{L}$, respectively (we fixed $z_{\rm d}=0.54$, $z_{\rm s}=1.49$, and $z_{\rm SNIa}=0.54$). The marginalised posterior distributions for the TDC, CC, and SNIa techniques are shown in blue, orange, and green, respectively. 
    The total marginalised posterior distributions (black) obtained by combining the TDC and CC, and TDC, CC, and SNIa methods are shown in the left and right panels, respectively. The 16th, 50th, and 84th percentiles of the combined marginalised distributions are highlighted with vertical dashed lines and the associated values are reported. The fiducial values are marked in red. The contour levels on the planes represent the 1, 2, and 3$\sigma$ confidence regions.}
    \label{fig:Bayesian-2params}
\end{figure*}

\section{Simulations}
\label{sec:sims}

In this section, we describe our simulation techniques exploiting two different approaches: one Bayesian and one Monte Carlo. As in Letter 1, we consider \macs1149, located at $z_{\rm d}=0.54$, as our reference lens cluster \citep{Grillo2016, Treu2016, Lotz2017}. The time-varying multiply-imaged source, a Type II SN known as SN `Refsdal', is at $z_{\rm s}=1.49$ \citep{Kelly2015, Kelly2016}.
By measuring the value of the Hubble parameter at the effective redshift of \macs1149 (see Sect.~\ref{sec:discussion}), we obtain $H(z)=H(z_{\rm d})$.

The Bayesian method assumes the following general form for the likelihood function associated with the two different probes $i$, namely, TDC and CC:
\be
\label{eq:lnLK_eq}
\mathcal{L}_i = \frac{1}{\sigma_i \sqrt{2\pi}}\exp(-\chi^2_i/2),
\ee
where 

\be
\label{eq:chi2}
\chi^2_{\rm TDC} = \left(\frac{D_{\Delta t, ~\rm true} - \hat{D}_{\Delta t}}{\sigma_{D_{\rm \Delta t, ~\rm true}}}\right)^2 \ \mathrm{and}\quad  
\chi^2_{\rm CC} = \left(\frac{H_\text{true}(z_{\rm d}) - \hat{H}(z_{\rm d})}{\sigma_{H_\text{true}(z_{\rm d})}}\right)^2,
\ee
for the TDC and CC methods, respectively. In these equations, $D_{\Delta t, ~\rm true}$ and $H_{\rm true}(z_{\rm d})$ are the values of the time-delay distance and of the Hubble parameter (at the redshift of the considered lens cluster, $z_{\rm d}$) computed in the fiducial input cosmological model that assumes $H_0 = 70~\mathrm{km~s^{-1}~Mpc^{-1}}$, $\Omega_{\rm m}=0.3$, $\Omega_{\rm de}=0.7$, and $w=-1$. The quantities $\hat{D}_{\Delta t}$ and $\hat{D}^\text{L}$ correspond to these distances obtained by sampling the two-dimensional ($H_0, ~\Omega_{\rm m}$) or four-dimensional ($H_0, ~\Omega_{\rm m}, ~ \Omega_{\rm de} , ~ w$) parameter space (considering $z_{\rm d}=0.54$ and $z_{\rm s}=1.49$).
Here, we adopt realistic errors on the relevant quantities $D_{\Delta t}$ and $H(z)$, namely, $\sigma_{D_{\Delta t,\text{true}}}$ and $\sigma_{H_\text{true}(z_{\rm d})}$, ranging from 5\% to 10\% and from 10\% to 20\%, respectively. In this work, we adopt these intervals to represent different scenarios. 
The total likelihood function accounting for the combination of the two techniques is obtained by multiplying the individual likelihood functions of each method as
\be
\label{eq:Ltot}
\mathcal{L}_\text{tot} = \mathcal{L}_{\rm TDC} \times \mathcal{L}_{\rm CC}.
\ee
In order to define the parameter posterior distributions, the following flat priors are assumed on the values of the cosmological parameters: $H_0 \in [20, 120]~\mathrm{km~s^{-1}~Mpc^{-1}}$, $\Omega_{\rm m}$ and $\Omega_{\rm de} \in [0, 1]$, and $w\in [-2, 0]$. 
The final log-posterior distributions are sampled by using \texttt{emcee}, which is the Python implementation of the affine-invariant Markov chain Monte Carlo (MCMC) ensemble sampler \citep{Goodman-Weare2010, emcee}. 
We performed our analysis adopting a flat-$\Lambda$CDM ($\Omega_{\rm m} + \Omega_{\rm de} = 1$ and $w=-1$) or an open-$w$CDM cosmological model.
The two- ($H_0$, $\Omega_\text{m}$) and four-dimensional ($H_0$, $\Omega_\text{m}$, $\Omega_\text{de}$, $w$) parameter space is then explored by using ten walkers, performing $10^5$ steps each. We note that the first 5000 steps of the walkers are removed as the burn-in phase, which corresponds to $\sim$100 times the integrated auto-correlation time of the parameters. 

Differently, in the Monte Carlo approach, we extracted $10^6$ realisations of $D_{\Delta t}$ and $H(z_{\rm d})$ from Gaussian distributions centred on the values of $D_{\Delta t, ~\rm true}$ and $H_\text{true}(z_{\rm d})$ (see above) and with standard deviations corresponding to the errors reported in Table~\ref{table:1}. Subsequently, by considering a two- ($H_0, ~\Omega_{\rm m}$) or a four-dimensional ($H_0, ~\Omega_{\rm m}, ~ \Omega_{\rm de} , ~ w$) grid divided into 1000 equally spaced bins and covering the assumed prior intervals, we computed the values of $D_{\Delta t}$ and $H(z_{\rm d})$ for each possible combination of cosmological parameters (corresponding to the different grid points). Finally, we determined the combination of cosmological parameter values that best reproduces each of the $10^6$ simulated measurements by computing the $\chi^2$ functions corresponding to the TDC and CC techniques, separately and combined (see Eq.~\ref{eq:chi2}). 

Both simulation methods yield median values of the cosmological parameters and associated precision in excellent agreement, proving the robustness of our analysis. Given that the Bayesian approach is more efficient and robust in sampling the posterior distribution, especially in a high-dimensional parameter space \citep{Wolz2012}, we present in the following only the results obtained with this technique.

\begin{table*}
  \renewcommand\arraystretch{1.4}
  \centering
  \caption{Intervals at the 68\% confidence level for the values of the cosmological parameters obtained with the Bayesian method.}
  \label{table:1}
  \begin{tabular}{cccccccc}
    \hline
    \hline
    & &
    \multicolumn{2}{c}{flat-$\Lambda$CDM$^{\mathrm{a}}$} &
    \multicolumn{4}{c}{open-$w$CDM} \\
    \cmidrule(r){3-4}\cmidrule(l){5-8}%\cmidrule(l){6-7}\cmidrule(l){8-9}
Err.$^\mathrm{b}$ $D_{\Delta t}$ & Err.$^\mathrm{b}$ $H(z_{\rm d})$ & $H_0$$^{\mathrm{c}}$ & $\Omega_{\rm m}$ & $H_0$$^{\mathrm{c}}$ & $\Omega_{\rm m}$ & $\Omega_{\rm de}$ & $w$ \\
    \hline
$5\%$ & $10\%$ &  $69.4^{+3.8}_{-3.6}$ &  $0.33^{+0.19}_{-0.15}$ &  $69.9^{+6.5}_{-5.5}$ &  $0.43^{+0.36}_{-0.31}$ &  $0.71^{+0.19}_{-0.29}$ &  $-1.18^{+0.55}_{-0.55}$ \\
$5\%$ & $20\%$ &  $69.0^{+3.9}_{-3.6}$ &  $0.40^{+0.31}_{-0.24}$ &  $69.8
^{+6.3}_{-5.3}$ &  $0.45^{+0.36}_{-0.31}$ &  $0.58^{+0.28}_{-0.34}$ &  $-1.14^{+0.65}_{-0.59}$ \\
$10\%$ & $10\%$ &  $68.9^{+7.5}_{-6.6}$ &  $0.33^{+0.24}_{-0.17}$ &  $68.1^{+9.3}_{-7.1}$ &  $0.41^{+0.36}_{-0.29}$ &  $0.65^{+0.23}_{-0.32}$ &  $-1.18^{+0.59}_{-0.56}$ \\
$10\%$ & $20\%$ &  $68.9^{+7.4}_{-6.3}$ &  $0.39^{+0.32}_{-0.24}$ &  $69.1^{+9.0}_{-7.2}$ &  $0.44^{+0.36}_{-0.31}$ &  $0.58^{+0.28}_{-0.35}$ & $-1.13^{+0.66}_{-0.60}$ \\  \hline
\hline
\end{tabular}
\begin{list}{}{}
\item[] 
$^{\mathrm{a}}$ $\Omega_{\rm m}+\Omega_{\rm de}=1$ and $w=-1$. \\
$^{\mathrm{b}}$ Adopted percentage relative errors. \\
$^{\mathrm{c}}$ ($\mathrm{km~s^{-1}~Mpc^{-1}}$).
\end{list}
\end{table*}

\section{Discussion}
\label{sec:discussion}

Table~\ref{table:1} summarises the median values and the 1$\sigma$ confidence intervals of the cosmological parameters within the adopted cosmological models and assuming different values for the uncertainties on $D_{\Delta t}$ and $H(z_{\rm d})$. 
In the left panel of Fig.~\ref{fig:Bayesian-2params}, we show, in a flat-$\Lambda$CDM model, the marginalised posterior probability distributions and the 1, 2, and 3$\sigma$ confidence regions for $H_0$ and $\Omega_{\rm m}$ inferred from the TDC (blue) and the CC (orange) techniques, and their combination (black), when assuming a 5\% and 10\% relative error for each technique, respectively. In the right panel of the same figure, we also assess the improvement of the precision on the values of these cosmological parameters from the combination of the TDC and CC techniques with an independent SN Ia luminosity distance ($D^\text{L}$) measurement in a cluster member galaxy, as newly proposed in Letter 1. This is done by including an additional $\mathcal{L}_{\rm SNIa}$ term (where the corresponding $\chi^2_{\rm SNIa}$ is defined in Eq.~(6) in Letter 1 and a 5\% relative uncertainty on the value of $D^\text{L}$ is assumed) in Eq.~(\ref{eq:Ltot}). 
We note that the uncertainty on $H_0$ is always driven by that of the term $D_{\Delta t}$. This is also shown by the fact that, for a fixed value of $\sigma_{D_{\Delta t}}$, considering more or less conservative relative errors on $H(z_{\rm d})$ (see Table~\ref{table:1}) or $D^\text{L}$ (see Letter 1) yields a similar precision on the measurement of $H_0$. When considering the more general open-$w$CDM model, its value is still robustly measured, with a slightly larger posterior probability distribution than in the flat-$\Lambda$CDM model. This result demonstrates that the estimate of $H_0$ is only mildly dependent on the assumed cosmological model. 

As discussed in Sect.~\ref{sec:methods} and in Letter 1, the three techniques are more sensitive to the value of $H_0$ than to those of the other cosmological parameters. For instance, the marginalised posterior probability distribution functions of $\Omega_{\rm m}$ are flat when considering each method alone. Remarkably, their combination allows for an estimate of the value of $\Omega_{\rm m}$ (with a significant statistical error, between $\sim 50 \%$ and $\sim 70 \%$ for more or less conservative scenarios; see Table~\ref{table:1}). 
This can be explained by the orientations of the $H_0$-$\Omega_{\rm m}$ intrinsic degeneracy from the different cosmological probes.
In particular, the $H_0$-$\Omega_{\rm m}$ degeneracies from the TDC and CC techniques are oriented in different directions, such that the overlapping region is smaller than for the joint TDC and SN~Ia method.
Thus, when considering the most optimistic scenario in the flat-$\Lambda$CDM model, the combined TDC and CC technique yields improved measurements of $H_0$ and $\Omega_{\rm m}$ by about a factor of $\sim 1.06$ and $\sim 1.22$, respectively, compared to the joint TDC and SN~Ia method (see Letter 1). 
However, since the degeneracy between $H_0$ and $\Omega_{\rm m}$ from the SN Ia method lies in between that from the TDC and CC methods, the gain in precision from the combination of the three probes is negligible.
We also highlight that the measured values of the cosmological parameters and, in particular, that of $H_0$, obtained by combining two or three of the discussed techniques (TDC, CC, and SNIa) are always accurately recovered (see Table~\ref{table:1}).
 
This is a conservative estimate of the value of $\Omega_{\rm m}$, since  (for simplicity) we are neglecting the contribution of the family-ratio term. This quantity can be probed when large samples of multiple images at different redshifts are observed, as is the case in lens galaxy clusters \citep[e.g.][]{Caminha2016, Richard2021, Bergamini2023a}. Thus, the values of $\Omega_{\rm m}$, $\Omega_{\rm de}$, and $w$ can also be measured 
\citep[][]{Jullo2010, Linder2011, Grillo2018, Caminha2022a}. 

After illustrating the possibility of enhancing and complementing the TDC technique with the combination of an estimate of $H(z_{\rm d})$ from exploiting the most massive cluster member galaxies as CCs, we discuss the observational feasibility of implementing the proposed joint method.
Multiple studies suggest that early-type cluster galaxies have assembled most of their stellar mass through massive star-formation events at early cosmic times \citep{Bernardi1998, Gobat2008, FerreMateu2014, Khullar2022}, with little star formation happening thereafter \citep{Treu2005, Thomas2005}. Such objects constitute thus a very homogeneous population, which is illustrated by the observation of a tight red sequence in the colour-magnitude diagram \citep{Faber1987, Gladders2000, DeLucia2007}, being already in place at high redshifts \citep[e.g.][]{Gobat2008, Menci2008, Lidman2008, Hilton2009, Cerulo2016, Strazzullo2016}. 
Galaxy clusters are thus unique environments in which to exploit the old and homogeneous population of massive, early-type galaxies as CCs \citep{Stern2010a, Stern2010b}.

Several lens galaxy clusters count with extensive, high-quality spectro-photometric data that can be readily exploited to select high-purity samples of CC tracers. 
In particular, high-resolution, deep, multi-band imaging of cluster fields, spanning the near-ultraviolet to the near-infrared \citep{Postman2012, Lotz2017, Estrada2023}, is valuable for distinguishing star-forming and passive populations \citep{Moresco2022}.
Thanks to its powerful capabilities, the Multi Unit Spectroscopic Explorer \citep[MUSE,][]{Bacon2012} integral-field spectrograph, on the Very Large Telescope (VLT), is the ideal instrument for carrying out spectroscopic studies of galaxies residing in galaxy cluster cores. 
With a fairly large field of view ($\sim$1 arcmin$^{2}$), MUSE has high spatial sampling (0.2$\arcsec/{\rm px}$), relatively high spectral resolution ($R\sim 3000$), and large wavelength coverage (4750 - 9350 $\AA$).
By targeting lens cluster fields, a large sample of homogeneous, deep (with typical exposure times of $\sim 5$ hours), and high-resolution spectra of cluster galaxies, over a wide redshift range $0.3<z<0.9$, has been collected \citep{Caminha2019, Bergamini2019, Richard2021, Jauzac2021, Lagattuta2022, Bergamini2023a, Bergamini2023b, Caminha2022b}. In addition, these MUSE observations allow for robust stellar velocity dispersion measurements, $\sigma_{\star}$, for statistically significant samples of member galaxies \citep{Bergamini2019, Pignataro2021, Granata2022, Bergamini2023a}. 

\begin{figure}[]
    \centering
    \includegraphics[width=\linewidth]{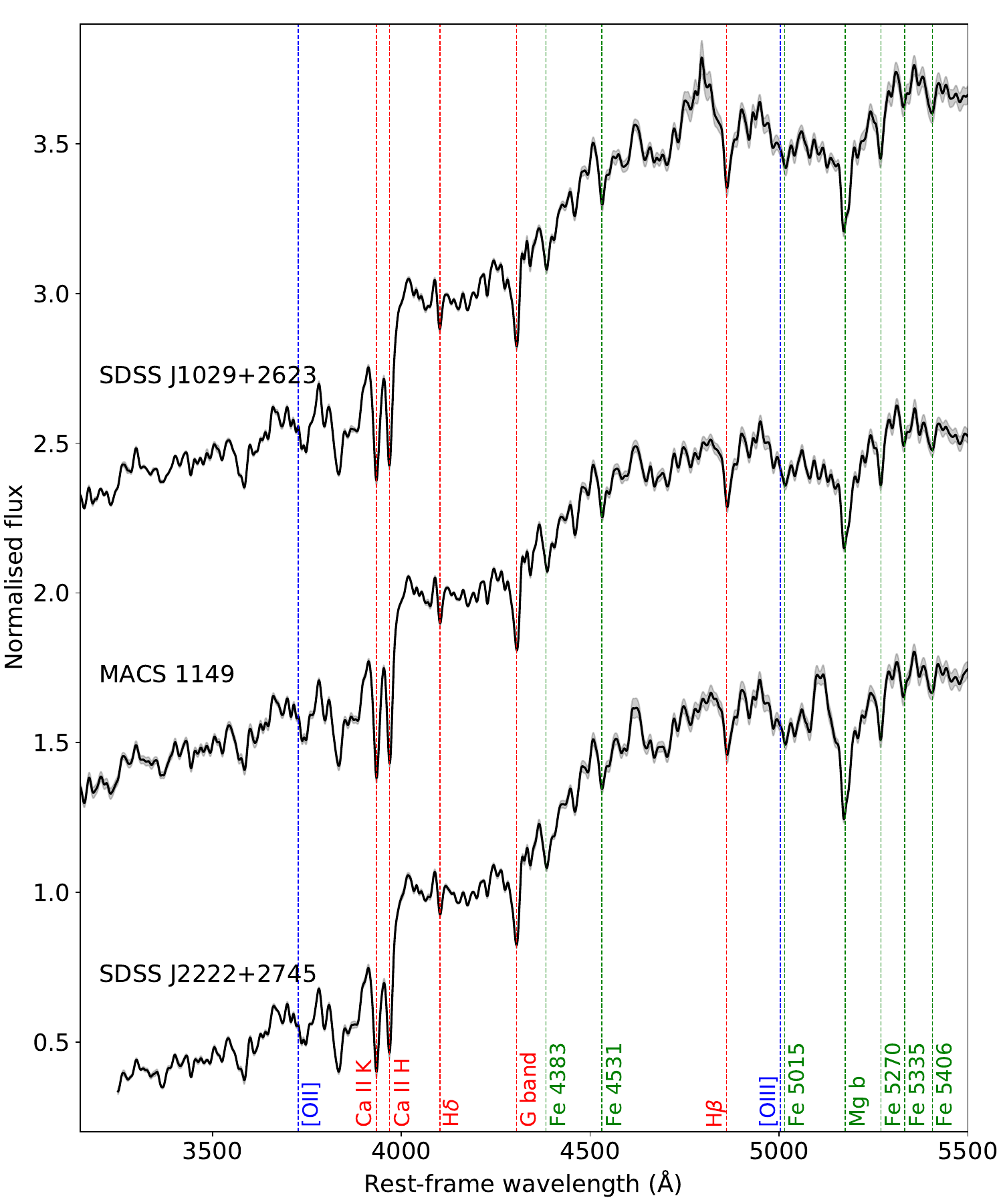}
    \caption{VLT/MUSE mean stacked spectra of 9, 22, and 8 cluster member galaxies of SDSS~J1029$+$2623 (top, $z=0.59$), \macs1149 (middle, $z=0.54$), and SDSS~J2222$+$2745 (bottom, $z = 0.49$), respectively, with $\sigma_{\star} > 180 ~{\rm km~s^{-1}}$. The shaded regions are the standard deviation of each spectral pixel. The spectra are smoothed by applying a Gaussian kernel with a standard deviation of $3.75 ~\AA$. For a clear visualisation, the flux values of \macs1149 and SDSS~J1029$+$2623 have been increased by $1$ and $2$, respectively.  
    The red dashed vertical lines locate the absorption features detected in the spectra, while the blue lines mark the expected positions of emission lines characteristic of young, star-forming populations. The green dashed lines denote the Fe and Mg lines typically used to estimate the stellar metallicity of the population. We note that the strong residual emissions observed at 4800 \AA\, and 5100 \AA\, for SDSS~J1029$+$2623 and SDSS~J2222$+$2745, respectively, are due to an incomplete subtraction of sky lines.\\
}
    \label{fig:Spectrum}
\end{figure}

For instance, \macs1149 has been observed with the VLT/MUSE instrument over an area of $\sim 2~{\rm arcmin}^2$ with an exposure time of $\sim 6$ hours \citep[see][Schuldt et al. in prep.]{Grillo2016}, resulting in the spectroscopic confirmation of 134 cluster member galaxies. 
In Fig.~\ref{fig:Spectrum}, we show a VLT/MUSE mean stacked spectrum of the 22 early-type cluster member galaxies with high stellar velocity dispersion values ($\sigma_{\star} > 180 ~{\rm km~s^{-1}}$) and characterised by a high signal-to-noise ratio (S/N $> 15$), showcasing the high S/N already achieved with a small sample. 
To create the stacked spectrum, we normalised the rest-frame shifted spectra of the selected galaxies by their median flux value in the rest-frame 4125 - 4275 $\AA$ wavelength range, where no prominent spectroscopic features are detected. The mean stacked spectrum has been smoothed with a Gaussian kernel with a standard deviation of $3.75~\AA$, allowing for a better visualisation of the spectral features.
The resulting spectrum is representative of an old, massive, and passively evolving galaxy population, as suggested by the red continuum, the absence of emission lines (their expected positions are identified in blue), and the presence of several absorption lines (marked in red) or spectral features (such as the 4000 $\AA$ break) that are commonly used to estimate their age. In addition, some key Fe and Mg absorption features \citep{Vazdekis2010}, marked in green, can help estimate the metallicity of the stellar population and, thus, control the intrinsic age-metallicity degeneracy \citep[see][]{Moresco2022}.
Finally, to measure the value of $\mathrm{d}t$ (see Eq.~\ref{eq:hubble_parameterII}) at the effective redshift of \macs1149 ($z_{\rm d}=0.54$), we can exploit already available VLT/MUSE observations of other lens clusters in close-by redshift bins such as SDSS~J1029$+$2623 ($z=0.59$) and SDSS~J2222$+$2745 ($z = 0.49$). Within the single VLT/MUSE pointings of these cluster cores \citep{Acebron2022b, Acebron2022}, 9 and 8 galaxies with $\sigma_{\star} > 180 ~{\rm km~s^{-1}}$ have been identified, respectively. Their resulting mean stacked spectra are also shown in Fig.~\ref{fig:Spectrum}.
These are characterised by a high average S/N, similar to that reported in previous studies \citep[][private communication]{Stern2010a, Moresco2016, Borghi2022b}. We plan to exploit the extensive, high-quality VLT/MUSE spectroscopic coverage already available for several lens clusters, enabling the selection of a high-purity sample of optimal CCs to ultimately apply the newly proposed joint method.

\section{Conclusions}
\label{sec:conclusions}

In this work, we present a novel technique to boost and complement the power of time-delay cosmography in lens galaxy clusters by probing their member galaxies as pure samples of CCs, allowing for an independent measurement of the value of the Hubble parameter at the effective lens cluster redshift.  

Considering as reference the well-studied lens galaxy cluster MACS~J1149.5$+$2223, with the multiply-imaged SN ‘Refsdal’, we have assessed the complementarity of the two probes and quantified the achieved precision on the values of the relevant cosmological parameters through their combination. 
We have demonstrated that the estimate of the value of the Hubble constant is robust, depending only mildly on the chosen cosmological model. Since the two probes produce confidence regions on the cosmological parameter planes that are oriented in complementary ways, we have shown that their combination provides more precise measurements of the values of the other cosmological parameters. 
Finally, we  discuss the immediate observational feasibility of the proposed joint method by exploiting the already available high-quality spectro-photometric data. These are necessary to select pure samples of cluster members as CCs and measure their ages for several cluster strong lensing systems.

Thanks to upcoming facilities such as the Legacy Survey of Space and Time, operated by the Vera C. Rubin Observatory, and such space missions as Euclid and JWST, a notable increase of known strong lensing clusters with time variable sources is expected. This will allow us to exploit them in combination with CCs as powerful cosmological probes. 

\begin{acknowledgements}
We kindly thank the anonymous referee for the constructive comments that helped to improve the clarity of the manuscript. We acknowledge financial support through grants PRIN-MIUR 2017WSCC32 and 2020SKSTHZ.
AA has received funding from the European Union’s Horizon 2020 research and innovation programme under the Marie Skłodowska-Curie grant agreement No 101024195 — ROSEAU.
SHS thanks the Max Planck Society for support through the Max Planck Fellowship.
This research was supported by the Munich Institute for Astro-, Particle and BioPhysics (MIAPbP) which is funded by the Deutsche Forschungsgemeinschaft (DFG, German Research Foundation) under Germany’s Excellence Strategy – EXC-2094 – 390783311.
This work uses the following software packages:
\href{https://github.com/astropy/astropy}{\texttt{Astropy}}
\citep{astropy1, astropy2},
\href{https://github.com/dfm/corner.py}{\texttt{Corner.py}}
\citep{corner},
\href{https://github.com/dfm/emcee}{\texttt{Emcee}}
\citep{emcee},
\href{https://github.com/matplotlib/matplotlib}{\texttt{matplotlib}}
\citep{matplotlib},
\href{https://github.com/numpy/numpy}{\texttt{NumPy}}
\citep{numpy1, numpy2},
\href{https://pypi.org/project/ppxf/}{\texttt{pPXF}}
\citep{Cappellari2023},
\href{https://www.python.org/}{\texttt{Python}}
\citep{python},
\href{https://github.com/scipy/scipy}{\texttt{Scipy}}
\citep{scipy}.
\end{acknowledgements}


%\begin{thebibliography}{}
\bibliographystyle{aa} % style aa.bst
\bibliography{refs} % your references Yourfile.bib
%\end{thebibliography}

\begin{appendix} %First appendix

\end{appendix}

\end{document}